\documentclass[aps,prb,reprint,groupedaddress,showpacs]{revtex4-1}
\usepackage{amsmath}
\usepackage{bm}
\usepackage{graphicx}
\usepackage{longtable}
\usepackage[force]{feynmp-auto}
\usepackage[colorlinks=true]{hyperref}
\RequirePackage{ifthen}
\allowdisplaybreaks
\usepackage[usenames,dvipsnames]{color}
\usepackage{ulem} 

\newcommand{\green}{\textcolor{OliveGreen}}

\begin{document}

\title{Dipolar phonons and electronic screening in monolayer FeSe on SrTiO$_3$}


\author{Yuanjun Zhou}
\affiliation{Department of Physics, Columbia University, New York, New York 10027, USA}
\author{Andrew J. Millis}
\affiliation{Department of Physics, Columbia University, New York, New York 10027, USA}


\date{\today}

\begin{abstract}
Monolayer films of FeSe grown on SrTiO$_3$ substrates exhibit  significantly higher superconducting transition temperatures than those of bulk FeSe. Interaction of electrons in the FeSe layer with dipolar SrTiO$_3$ phonons has been suggested as the cause of the enhanced transition temperature.  In this paper we systematically study the coupling of SrTiO$_3$ longitudinal optical phonons to the FeSe electron, including also electron-electron Coulomb interactions at the  random phase approximation level. We find that the electron-phonon interaction between FeSe and SrTiO$_3$ substrate is almost  entirely  screened by the electronic  fluctuations in the FeSe monolayer, so that the net electron-phonon interaction is very weak and unlikely to lead to superconductivity. 
\end{abstract}



\maketitle

\def\scr{\scriptsize}
\def\draftversion{false}
\def\draftversion{true}
\ifthenelse{\equal{\draftversion}{true}}{
  \marginparwidth 2.7in
  \marginparsep 0.5in
  \newcounter{comm} 
  \def\commnext{\stepcounter{comm}}
  \def\commtext{{\bf\color{blue}[\arabic{comm}]}}
  \def\commmar{{\bf\color{blue}[\arabic{comm}]}}
  \def\yjm#1{\commnext\marginpar{\small YJZ\commmar: #1}\commtext}
  \def\ajm#1{\commnext\marginpar{\small AJM\commmar: #1}\commtext}
  \def\sm#1{\commnext\marginpar{\scr\green{SAVED\commmar: #1}}\commtext}
  \def\tnewpage{\marginpar{\small Temporary newpage}\newpage}
}{
  \def\yjm#1{}
  \def\ajm#1{}
  \def\sm#1{}
  \def\tnewpage{}
}

\section{Introduction}
The recent discovery of superconductivity, or its signatures at temperatures of orders 70K (tunneling measurements) and 100K({\it in situ} four-point measurements) in monolayers of FeSe grown on the (001) and (111) surfaces of niobium-doped SrTiO$_3$(STO)\cite{FeSe-CPL,He2013,Ge-100K} challenges our understanding of superconductivity in the pnictide compounds and has stimulated intense research activity. Monolayer FeSe on STO is heavily electron doped relative to bulk FeSe\cite{Tan-nmat}. Surface potassium doping of free-standing  FeSe films \cite{FeSe-Kdoped,WHZhang16} produces transition temperatures as high as 45K, and systematic variation of carrier concentration using gate doping with liquid dielectrics reveals that the high transition temperature appears at the point where the doping is large enough to eliminate the zone center hole pockets\cite{FeSe-flake-gating}.  However, the highest transition temperatures induced  by  pure electron doping are about 45K,  still notably less than the 70K or 100K\cite{Ge-100K} reported  for  monolayer FeSe on STO, and recent studies of  monolayer FeSe on anatase TiO$_2$ report similarly high transition temperatures \cite{Ding-FeSeTiO2}, strongly suggesting that an additional substrate-specific $T_c$ enhancement occurs. One clue to the nature of the substrate-specific interactions is provided by recent angle-resolved photoemission spectroscopy (ARPES) measurements\cite{el-ph_Shen-FSSTO}, which reveal ``replica'' bands, images of the FeSe conduction band shifted up in binding energy by an amount comparable to one of the longitudinal optical (LO) phonon energies in SrTiO$_3$. These bands are attributed to interaction of electrons in the FeSe with very long wavelength optical phonons in the SrTiO$_3$.

Theoretical papers have appeared analyzing the interfacial enhancement of the $T_c$ and the presence of replica bands in monolayer FeSe on STO via the electron -LO phonon interaction\cite{DHLee15-CPB,Rademaker16}.  However, these studies focus mainly on the electron phonon interaction, neglecting the  Coulomb interaction between the FeSe electrons. This interaction may screen the electron-phonon interaction. In an extreme antiadiabatic limit and with Thomas-Fermi screening, Gor'kov\cite{Gorkov16} recently studied the electron phonon interactions in the FeSe/STO system and argued that the LO phonons from STO are not enough to induce such a high $T_c$ found in the FeSe/STO system.

In this paper, we present an analysis that treats the STO LO phonons and the FeSe Coulomb interaction on an equal footing.  We find that although the LO phonons in STO generates an attractive potential, it is strongly screened by the electrons in FeSe layer, so that the electron-phonon interaction suppressed, producing neither replica bands nor an appreciable contribution to superconductivity for reasonable parameters.
This is similar to the result found in Inkson and Anderson on plasmon-mediated superconductivity.\cite{Inkson73}.

The paper is organized as follows. Section~\ref{sec.el-ph} presents the screened  electron phonon interaction in FeSe/STO. 
Section~\ref{sec.RPA} gives a RPA level analysis of the total interaction.
Section~\ref{sec.Plasmon} discusses the 2D plasmons in FeSe. 
Section~\ref{sec.interaction} analyses the net phonon contribution to the electron-electron interaction.
 In section~\ref{sec.replica} we discuss the possibility of replica bands.
Finally Section~\ref{sec.summary} is a summary and conclusion. 
Appendices give details of derivations.

\section{Electron-phonon interaction}
\label{sec.el-ph}
We consider a monolayer of FeSe grown on top of a semi-infinite SrTiO$_3$ crystal. We take the STO to occupy the $z<0$ half-plane and assume that the electrons in the FeSe occupy a layer of negligible thickness at a distance $z_1 > 0$ from the surface. In the actual system, $z_1\approx 4.3$\AA.  In the SrTiO$_3$ we consider that each unit cell $i$ hosts several atomic displacement modes, labelled by an index $a$ and characterized by a displacement vector $\vec{d}_i^a$  with  effective charge $Z_ae$ (e is the electron charge) so the dipole moment due to a given ionic displacement is $Z_a\vec{d}_i^a$. We Fourier transform on the in-plane coordinates and label the planes parallel to the interface by $J$ so that the dipole moment is $Z_a\vec{d}(q)_J^a$. In FeSe we focus on the electronic charge density $-e\rho$ which we write as a function of the in-plane momentum $q$.

To derive the Hamiltonian we write the total Coulomb energy as 
\begin{eqnarray}
H_{Coul}&=&\int \frac{d^2q}{(2\pi)^2}\mathcal{H}_{DD}(q)+\mathcal{H}_{D\rho}(q)+\mathcal{H}_{\rho\rho}(q)
\end{eqnarray}
with
\begin{eqnarray}
\mathcal{H}_{DD}(q)&=&\frac{1}{2}\sum_{JJ^\prime ab}V_{DD}(J,J^\prime,q)Z_aZ_bd(q)_J^ad(-q)_{J^\prime}^b
\\
\mathcal{H}_{D\rho}(q)&=&\sum_{Ja}V_{D\rho}(J,q)Z_ad(q)_J^a\rho(-q)
\\
\mathcal{H}_{\rho\rho}(q)&=&\frac{1}{2}V_{\rho\rho}(q)\rho(q)\rho(-q)
\end{eqnarray}
where $V_{DD}(q,(J,J^\prime,q)$ gives the interaction energy between dipoles of unit charge and in-plane momentum $q$ in layers $J,J^\prime$, etc. 

\begin{figure}[ht]
 \includegraphics[width=\columnwidth]{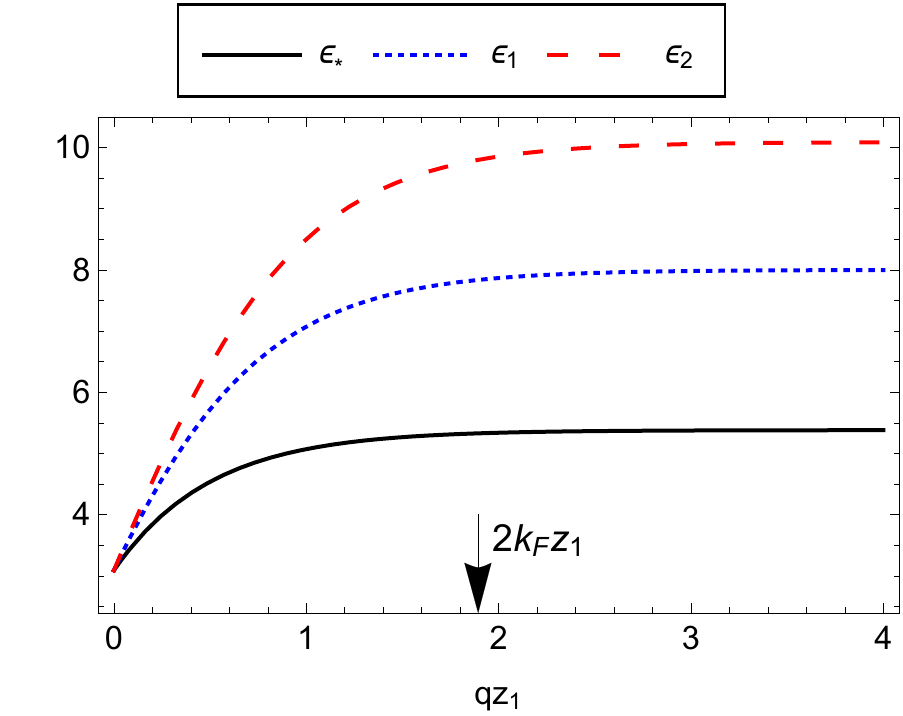}%
 \caption{\label{fig.eps} High frequency (static) dielectric constants, $\epsilon_\star$, $\epsilon_1$ and $\epsilon_2$ with respect to wavevector. We take $\epsilon_\text{STO} = 5.18$, and $\epsilon_\text{FeSe}=15$.}
  \end{figure}

Determining the interactions $V$   requires solving an electrostatics problem  that is complicated by the spatial asymmetry (vacuum at $z>z_1$ and SrTiO$_3$ at $z<0$) and the lack of momentum conservation in the $z$ direction. However, useful simplifications occur in the long wavelength limit of interest here. Details are given in Appendix~\ref{A.electrostatics}.
The results depend on three effective dielectric constants: $\epsilon_1$ parametrizing the strength of an electric field in the FeSe layer due to charges in this layer, $\epsilon_2$ parametrizing the strength of an electric field in the STO due to charges in the STO and $\epsilon_\star$ describing the field in the FeSe layer due to a charge in the STO. 
The dependence of the effective dielectric constant on wavevector is shown in Fig.~\ref{fig.eps} (this dependence is not denoted explicitly in the formulas that follow.)  Important parameters of $\epsilon_1$, $\epsilon_2$ and $\epsilon_\star$ are $\epsilon_\text{STO}$, and $\epsilon_\text{FeSe}$. $\epsilon_\text{STO}$ is  the ``high frequency'' dielectric constant of the STO,  formally defined as the bulk STO dielectric constant at frequencies much less that the SrTiO$_3$ band gap, if the dipolar phonons are frozen. Similarly, $\epsilon_\text{FeSe}$ is the dielectric constant of FeSe if the electronic charge fluctuations are frozen. Detailed formulas are in the Appendix~\ref{A.electrostatics} and the asymptotics are in Table~\ref{table.dielectric} (note in particular the simplicity of the long wavelength limit, in which half of the field lines are in vacuum and half in the STO). Note that the high frequency dielectric constants $\epsilon_\text{FeSe}$ and $\epsilon_\text{STO}$ are both small\cite{Yuan13_FeSe_refl, STO-T-permittivity, STO-field-permittivity} and the intrinsic momentum dependences of $\epsilon_1$, $\epsilon_2$ and $\epsilon_\star$ are  weak.

\begin{table}[ht]
\caption{\label{table.dielectric} Asymptotics of dielectric constants. }
 \begin{ruledtabular}
\begin{tabular}{c c c}
& $qz_1 \sim 0$  & $qz_1 \sim \infty$\\
\hline
$\epsilon_1$ & $(\epsilon_\text{STO} + 1)/2$ & $(\epsilon_\text{FeSe} + 1)/2$  \\
$\epsilon_2$ & $(\epsilon_\text{STO} + 1)/2$ & $(\epsilon_\text{STO} +  \epsilon_\text{FeSe})/2$ \\
$\epsilon_\star$& $(\epsilon_\text{STO} + 1)/2$ & $(\epsilon_\text{STO}+ \epsilon_\text{FeSe}) (\epsilon_\text{FeSe} +1) /4\epsilon_\text{FeSe}$ 
\end{tabular}
 \end{ruledtabular}
\end{table}

\begin{figure}[ht]
 \includegraphics[width=\columnwidth]{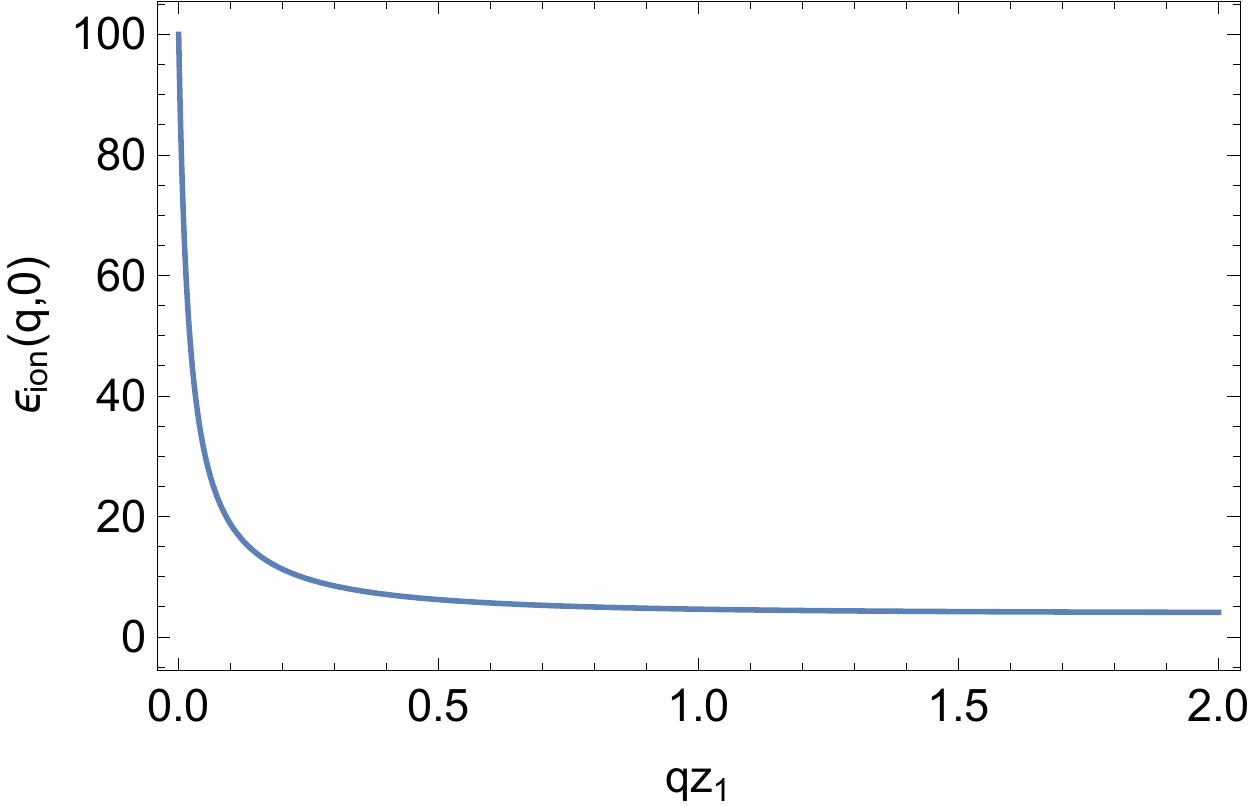}%
 \caption{\label{Fig.epsilon} $\epsilon_{ion}(q,\Omega)$ for zero frequency calculated for $2k_Fz_1 = 1.89$ representative of FeSe on STO. }
  \end{figure}

After solving the electrostatic problem and integrating out the phonons (see Appendix~\ref{A.decouple}) we obtain a Lagrangian describing the electron density fluctuations screened by the STO longitudinal phonons as 
\begin{eqnarray}
\label{Helfinal}
\mathcal{L}_{el}&=&\frac{1}{2}\int \frac{d^2q}{(2\pi)^2}\rho_q\rho_{-q}V_\text{eff}(q,\Omega)
\end{eqnarray}
with  effective electron-electron interaction
\begin{eqnarray}
V_\text{eff}(q,\Omega)&=&\frac{2\pi e^2}{q\epsilon_{ion}(q,\Omega)}
\label{Eq.Veff}
\end{eqnarray}
The ionic contribution to the dielectric function  is well approximated by
\begin{eqnarray}
\epsilon_{ion}(q,\Omega)&=&\epsilon_1 \left[ 1- e^{-2qz_1} \sum_a\frac{\gamma_a\Omega_a^2}{\Omega^2+\Omega^2_a} \right]^{-1} 
\label{Eq.epsilonion}
\end{eqnarray}
where the $\Omega_a$ are the frequencies of the longitudinal optic modes of SrTiO$_3$ as appropriately modified by the presence of the surface. The shifts of phonon frequencies due to the surface, the effect of the internal electric fields arising from the doped FeSe and the corresponding depletion region of the STO, and the other dielectric effects associated with spatial symmetry breaking are parametrized by  $\epsilon_2$ and $\epsilon_\star$ are included into the mode-dependent parameter $\gamma_a$ . The parameters $\gamma_a=0.002, 0.104, 0.854$ for $a = 1,2,3$ are chosen to be consistent with bulk STO and to produce a dielectric constant that  to coincides with our previous work\cite{yjzhou16}, in which $\epsilon(0,0)\approx 100$ was obtained for the near-interface region. 
The key feature of this result is the exponential dependence of the screening on momentum, shown in Fig.~\ref{Fig.epsilon} for parameters representative of FeSe on STO.

\section{Electronic Density Fluctuations: RPA screening and final effective interaction}
\label{sec.RPA}

The interaction in Eq.~\ref{Eq.Veff} is screened by electronic density fluctuations, which we treat here at the random phase approximation (RPA) level, leading to the final interaction
 \begin{equation}
V^\star(q,\Omega) = \frac{V_\text{eff}(q,\Omega)}{1-\chi_0(q,\Omega) V_\text{eff}(q,\Omega)}
\label{Eq.Vstar}
\end{equation}
with $\chi_0$ the free electron approximation to the electronic polarizability. 
We calculate $\chi_0$ in two dimensions and Matsubara frequency  assuming two parabolic bands representing the two zone-face-centered electron bands of FeSe:
\begin{equation}
\chi_0(q,\Omega)=4\int\frac{k dk d\theta}{4\pi^2}\frac{
f(\varepsilon_{k-\frac{q}{2}})-f(\varepsilon_{k+\frac{q}{2}})
}{
i\Omega+\varepsilon_{k-\frac{q}{2}}-\varepsilon_{k+\frac{q}{2}}\
}\equiv -N_0\Phi(q,\Omega)
\label{chi0def}
\end{equation}
with the prefactor 4 expressing the spin and valley degeneracy of the electrons and total density of states $N_0=2m/\pi$ with $m$ the electron mass.

It is  convenient to work in dimensionless units, defining 
\begin{eqnarray}
\bar{q}&=&\frac{q}{2k_F}
\\
\bar{z}&=&4k_Fz_1
\\
\bar{\Omega}&=&\frac{\Omega}{4E_F}
\end{eqnarray} 
with $E_F=k_F^2/(2m)$. We also introduce the gas parameter $r_S= 2/a_B^\star k_F$ where $a_B^\star = \epsilon_1/me^2$ is the effective Bohr radius.  In our previous work\cite{yjzhou16}, we found for band parameters appropriate to monolayer FeSe on SrTiO$_3$  $m\approx m_e$, and $k_F\approx 0.22/\AA$ so $E_F\approx 0.1eV$ (strong correlation effects may reduce this value).  With $\epsilon_1\approx 3$ for small wavevectors,  we have $a^\star_B \approx 3 a_B$, and so $r_S\approx 6$. We also rescale all dielectric functions by $\epsilon_1$ and denote the rescaled functions with tildes (e.g. $\tilde{\epsilon}_{ion}=\frac{\epsilon_{ion}}{\epsilon_1}$), so 

We then  define a dimensionless interaction $v^\star=N_0V^\star$ and obtain
\begin{equation}
v^\star(q,\Omega) =\frac{\frac{r_S}{\bar{q}{\tilde \epsilon}_{ion}}}{1+\frac{r_S}{\bar{q}{\tilde \epsilon}_{ion}}\Phi}=\frac{r_S}{\bar{q}\tilde{\epsilon}_{tot}}
\label{Vstardimensionless}
\end{equation}
with 
\begin{eqnarray}
\tilde{\epsilon}_{tot}&=&\tilde{\epsilon}_\text{RPA}+{\tilde \epsilon}_{ion}-1
\\
\tilde{\epsilon}_\text{RPA}&=&1+\frac{r_S}{\bar{q}}\Phi
\label{epsilonRPA}
\end{eqnarray}

\section{Plasmon}
\label{sec.Plasmon}
Plasmon frequencies are zeros of the total dielectric function, in other words frequencies $\omega_{pl}(q)$ satisfying
\begin{equation}
\epsilon_{tot}(q,\omega_{pl}(q))=0
\label{plasmon}
\end{equation}
Using Eqs.~\ref{Eq.epsilonion} and \ref{epsilonRPA} and adopting the small-q limit form for $\Phi$  we obtain for the plasmon dispersion
\begin{equation}
\frac{r_S}{\bar{q}}\left(1-\frac{\bar{\omega}_{pl}}{\sqrt{\bar{\omega}_{pl}^2-\bar{q}^2}}\right)+\left[1-\gamma e^{-\bar{q}\bar{z}}\sum_a\frac{\bar{\Omega}_a^2}{\bar{\Omega}_a^2-\bar{\omega}_{pl}^2}\right]^{-1}=0
\label{plasmon2}
\end{equation}
Note that for $\bar{q}>\bar{\omega}_{pl}$ the plasmon enters the particle-hole continuum and becomes overdamped and Eq.~\ref{plasmon2}, which gives the dispersion for undamped plasmons, does not apply. 

Let us first consider the limit $\bar{\omega}_{pl}\ll \bar{\Omega}_a$, in which case $\bar{q}\ll\bar{\omega}_{pl}$,  the last term is approximately $\left(\tilde{\epsilon}_{ion}(0,0)^{-1}+\bar{q}\bar{z}\right)^{-1}$ with $\tilde{\epsilon}_{ion}(0,0)\gg1$ and we have, approximately

\begin{equation}
\tilde\omega_{pl}^2=\frac{r_S\bar{q}}{2\tilde{\epsilon}_{ion}(0,0)}+\frac{r_S\bar{z}}{2}\bar{q}^2
\label{plasmon3}
\end{equation}

Thus at very low energies and long wavelengths we have a conventional 2D plasmon with square root dispersion determined by the long wavelength dielectric constant, which is large. At the scale $\bar{q}_z\approx 1/( \tilde{\epsilon}_{ion}(0,0)\bar{z})$ the wavevector dependence of the dielectric function (arising from the set-back of the FeSe layer from the SrTiO$_3$) becomes important and the dispersion becomes linear in wavevector. However, when $\omega_{pl}$ becomes comparable to the lowest optic phonon frequency, the frequency dependence of the dielectric function becomes important and the dispersion changes. 

Band theory for the monolayer FeSe on STO \cite{yjzhou16} gives  $r_S = 5.7$, $\bar z = 3.8$, $\tilde\epsilon(0,0)^{-1} = 0.04$. The lowest important phonon frequency $\bar\Omega_1 \approx 0.025$ so the momentum at which the plasmon crosses the phonon is about $\bar{q}\sim 0.005$; at this scale the quadratic term in Eq.~\ref{plasmon3} is small so the regime of linearly dispersing plasmons is obscured by the phonon bands. 

\begin{figure}[ht]
 \includegraphics[width=\columnwidth]{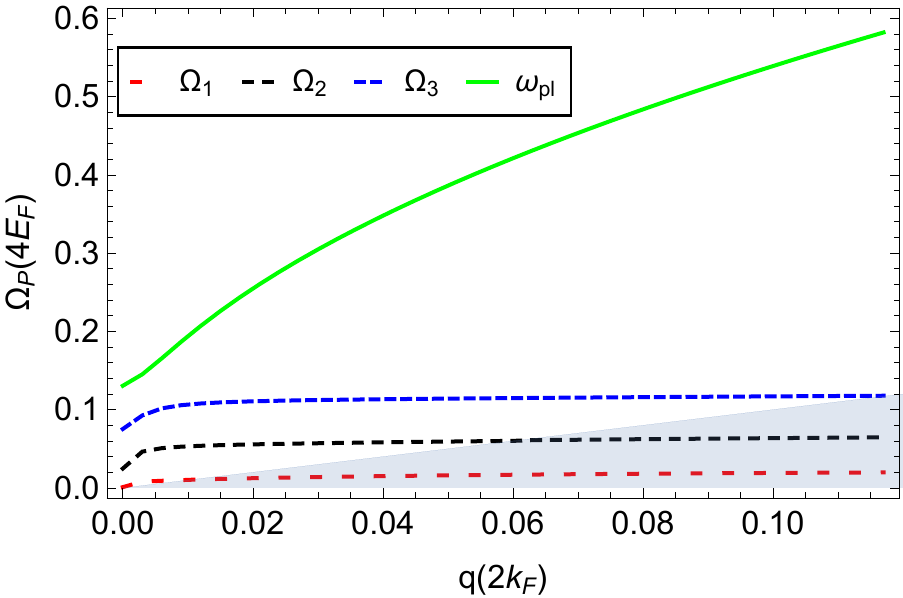}%
 \caption{\label{Fig.plasmon} Collective mode dispersions with the interaction with the STO LO modes for $r_S = 5.7$, $\bar z = 4k_Fz_1 = 3.89$. The shaded region represents the particle-hole continuum.}
\end{figure}

At frequencies above the highest phonon frequency the plasmon dispersion crosses over to the standard unrenormalized square-root dispersion. Compared to the long wavelength case, the only difference is the dielectric constant, which enters the frequency as the square root, so the dispersion curve is changed by a factor of about $4$. Fig.~\ref{Fig.plasmon} shows the locus of the zeros of the dielectric function for the band parameters obtained in our previous work \cite{yjzhou16}. The renormalized plasmon is visible only at the very lowest frequencies.

\section{Interaction \label{sec.interaction}}

There are various ways to view the combined electron-electron and electron-ion interactions. For a first approach we follow the  theory of electron-phonon interactions in semiconductors \cite{Varga65,Cohen67} and define the electron-phonon interaction as the difference between the total interaction and the purely electronic (here RPA) parts
\begin{equation}
v^\star(q,\Omega)=v_\text{RPA}(q,\Omega)-v_{ph}(q,\Omega)
\label{vstar2}
\end{equation} 
where 
\begin{equation}
v_\text{RPA}=\frac{\frac{r_S}{\bar{q}}}{1+\frac{r_S}{\bar{q}}\Phi}=\frac{1}{\frac{\bar{q}}{r_S}+\Phi}\equiv\frac{r_S}{\bar{q}\tilde{\epsilon}_\text{RPA}}
\label{vrpadef}
\end{equation} 
and (note the standard sign convention for which a positive $v_{ph}$ is an attractive interaction)
\begin{equation}
v_{ph}=v_\text{RPA}\frac{\tilde{\epsilon}_{tot}-\tilde{\epsilon}_\text{RPA}}{\tilde{\epsilon}_{tot}}=v_\text{RPA}\frac{\frac{\bar{q}}{r_S}\left(1-\tilde{\epsilon}_{ion}^{-1}\right)}{\frac{\bar{q}}{r_S}+\tilde{\epsilon}_{ion}^{-1}\Phi}
\label{vph}
\end{equation}

Refs.~\onlinecite{Varga65,Cohen67} argued that the $v_\text{RPA}$ term should be viewed as the RPA-screened electron-phonon coupling. However, in the present case the imaginary part of $v_{ph}$ changes sign as the frequency is varied, it cannot be interpreted as the longitudinal phonon propagator renormalized by coupling to electrons (see Appendix~\ref{A.convention} for details). 

In a conventional metal, $\tilde{\epsilon}$ has negligible momentum dependence, and is different from unity only for frequencies less than the longitudinal optic phonon frequency. For these frequencies, for all momenta except for the very narrow range $q<\Omega_{LO}/v_F$ we may set $\Phi=1$. For typical $\tilde{\epsilon}$ values somewhat larger than 1 and typical metallic $r_S \sim 2$ we find a dimensionless interaction of the order of (but somewhat smaller than) unity. Thus in many conventional superconductors longitudinal modes make some contribution to pairing but the transverse modes, which are not screened, make a larger contribution. 

To understand the differences arising in the present situation it is useful to consider a simplified situation in which there is only one optic phonon mode (bare frequency $\Omega_{LO}$) and assume that the host material is tuned exactly to the ferroelectric instability $\gamma=1$.  Then we may rewrite Eq.~\ref{vph} using Eq.~\ref{Eq.epsilonion} as (in the denominator we approximated $1-e^{-\bar{q}\bar{z}}\rightarrow \bar{q}\bar{z}$)

\begin{equation}
v_\text{ph} = v_\text{RPA}\frac{e^{-\bar{q}\bar{z}}\bar{\Omega}_{LO}^2}{\bar{\Omega}^2\left(1+\frac{r_S}{\bar{q}}\Phi\right)+\bar{\Omega}_{LO}^2
\left(1+r_S\bar{z}\Phi
\right)
}
\end{equation}

\begin{figure}[ht]
 \includegraphics[width=\columnwidth]{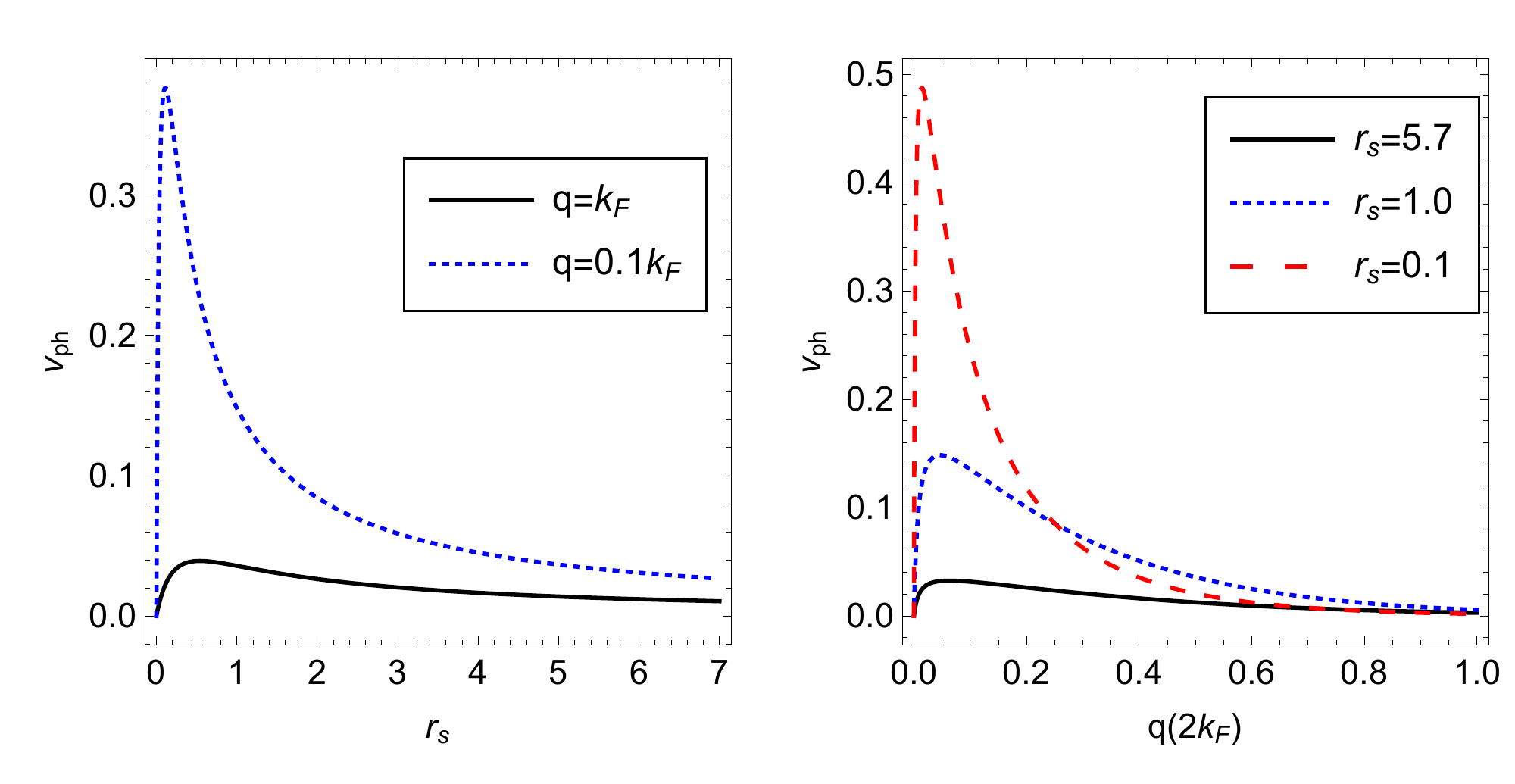}%
 \caption{\label{Fig.Vphrs} Zero frequency $v_\text{ph}$, left: as a function of $r_S$, and right: as a function of $q$. }
\end{figure}

As expected, the interaction is confined to momenta less than or of order of $\bar{z}^{-1}\approx 0.25$.  The frequencies $\bar{\Omega}=\bar{q}$ corresponding to these momenta are  of the order of the highest phonon frequency. Thus on the Matsubara axis, for $ \bar\Omega \lesssim \bar{q}$, $\Phi$ and $v_\text{RPA} \approx 1$ and the phonon contribution of the interaction $v_{ph}$ is approximately $1/r_S\bar z\lesssim 0.04$ using the band parameters mentioned above. These qualitative conclusions are confirmed by the two panels of Fig.~\ref{Fig.Vphrs} which show the $r_S$ and momentum dependence of $v_\text{ph}$ at zero Matsubara frequency.   We see that the interaction is enhanced at small momentum and small $r_S$ but is never even as large as unity.  

\begin{figure*}[t]
 \includegraphics[width=2\columnwidth]{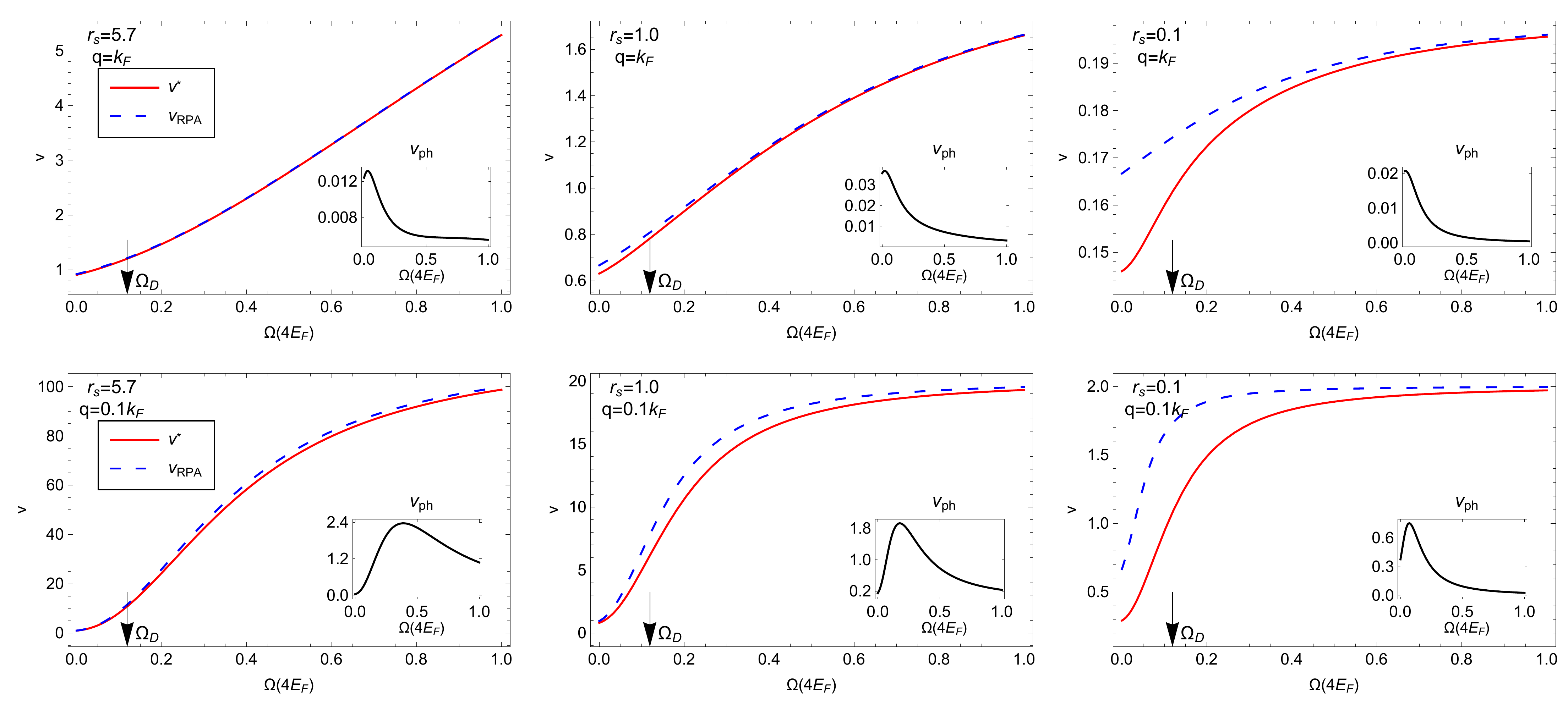}%
 \caption{\label{Fig.Vrs} Dimensionless interaction plotted as function of   frequency for $q$ and $r_S$-values shown. The insets show the relevant phonon-induced attractive interactions $v_\text{ph}$. $\Omega_D$, the highest and dominant LO phonon frequency, label the arrows on the frequency axes.}
\end{figure*}

For larger $\bar\Omega$ and smaller $\bar q$, $\Phi$ becomes small, $v_\text{RPA}$ becomes larger while the percentage of the $v_{ph}$ in the total interaction is reduced.  To obtain a significant interaction one must achieve a much smaller set-back distance, so that $\tilde{\epsilon}$ is large even for $\bar{q}\sim 1$  and have  more weakly correlated electron gas (smaller $r_S$). This qualitative analysis is confirmed by the detailed numerics presented in Fig.~\ref{Fig.Vrs}.  Thus in effect electronic screening strongly reduces the interaction, so very little significant effect of phonons on electronic properties remains. 

\section{Replica bands}
\label{sec.replica}
Recent photoemission experiments have reported replica bands in monolayer FeSe on STO  \cite{el-ph_Shen-FSSTO}. A replica band is an image of the main band, shifted to a higher binding energy by a momentum independent energy. Replica bands are obtained theoretically in electron-boson calculations involving an extreme forward-scattering limit \cite{Rademaker16} and it was suggested that coupling to STO dipolar phonons could satisfy the necessary conditions.

Here for simplicity we discuss the replica bands by calculating the electron self energy  to leading order in the interaction:
\begin{equation}
\Sigma(k,\omega)  = \int d\nu \frac{d^2q}{(2\pi)^3} G^0(k+q,\omega+\nu)V^\star(q,\nu) 
\label{Eq.sig1}
\end{equation}
with $G^0$ the free electron Green's function. In the extreme forward scattering limit and on the Matsubara axis
\begin{equation}
V^\star_{forward}=g_0^2\delta^2(q)\frac{2\Omega_0}{\nu^2+\Omega_0^2}
\label{forward}
\end{equation}
so analytically continuing
\begin{equation}
\Sigma_{forward}(k,\omega)  =\frac{g_0^2 n_f}{\omega-(\varepsilon_k-\Omega_0)-i\delta} + 
\frac{g_0^2 (1-n_f)}{\omega-(\varepsilon_k+\Omega_0)-i\delta}
\label{Eq.sig1forward}
\end{equation}
with $n_f$ the Fermi function.

\begin{figure}[t]
 \includegraphics[width=\columnwidth]{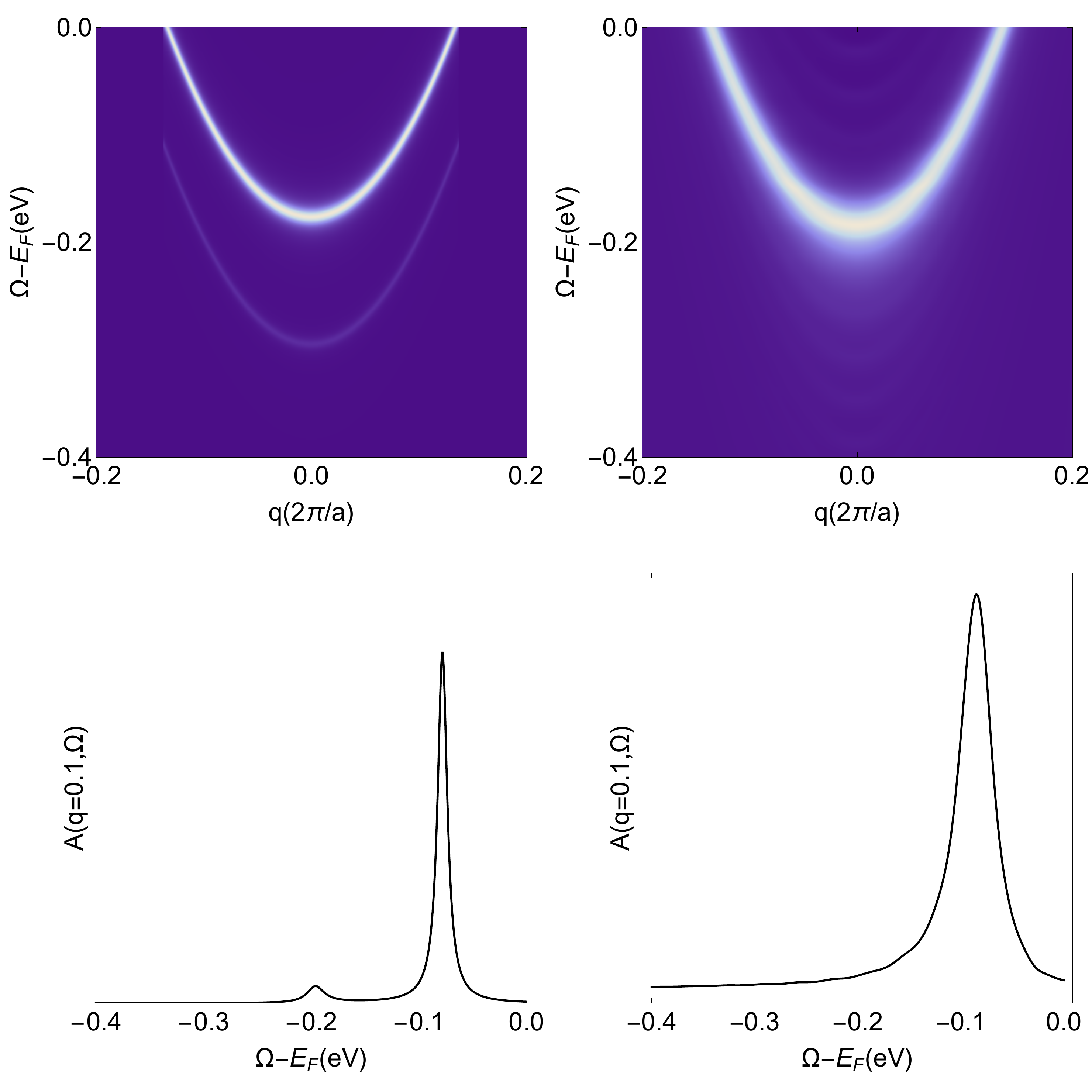}%
 \caption{\label{Fig.spect} Electron spectral function computed for idealized delta-function forward scattering model (Eq. ~\ref{Eq.sig1forward} (left panels) and physical electron-phonon coupling (right panels). Top panels: greyscale (color online) representation of spectral function as function of frequency and of  wavevector  along $\Gamma - M - \Gamma$. Bottom panels: energy dispersion curves computed at $q=0.1 (2\pi/a)$ away from M point. The self energy is calculated from Eq.~\ref{Eq.sig1} with the frequency integration cut at $|\omega'| = 20$ eV and $64\times 64$ q points within the Fermi surface. Here $a$ is the lattice constant of the physical (2Fe) Brillouin zone.}
  \end{figure}

In the upper and lower  left panels of Fig.~\ref{Fig.spect} we show the spectral function computed using Eq. ~\ref{Eq.sig1forward} for a line through the M point (center of the electron pocket). The main peak is the quasiparticle energy, and the replica band is visible at higher binding energy. The distance between the main peak and the shakeoff peak is dependent on the electron-phonon interaction matrix\cite{Rademaker16}.

\begin{figure}[t]
 \includegraphics[width=\columnwidth]{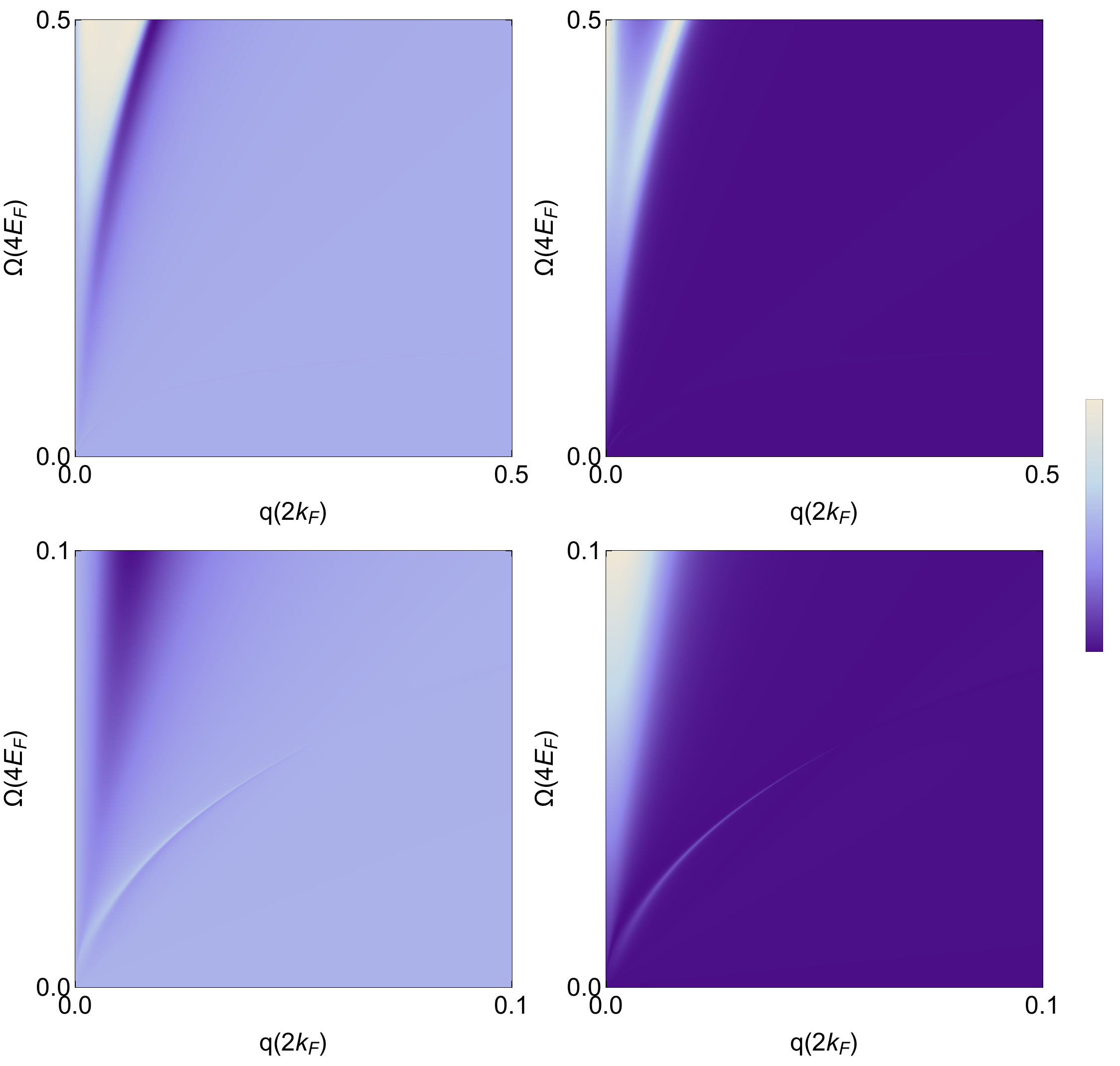}%
 \caption{\label{Fig.vreal} Greyscale plot of magnitude of  dimensionless interaction  $v^\star$  computed in plane of momentum and real frequency using  the simplified model containing only the  the highest LO phonon at  $r_S = 5.7$. Left, real part. Right: imaginary part. Bottom panels are the zoomed in parts of the upper panels for small wavevectors.}
\end{figure}  
 
Fig.~\ref{Fig.vreal}  shows the low frequency behavior of the real and imaginary parts of total interaction computed for FeSe/STO, computed using a simplified model in which only the highest frequency phonon is retained. We see that the interaction has considerable structure as a function of momentum and energy, and is not particularly peaked at small momentum.  For this reason, the interactions (which are essentially the same as in the absence of phonons), produce a wide tail in the energy dispersion curve  but no replica bands. 
 
Indeed, replica bands are not always reported in experiments monolayer FeSe on STO with superconducting $T_c$ greater than 60K, the replica bands are not always reported. Thus it is possible that certain surface treatments change the surface electronic states. Our calculation shows that replica bands are not a natural consequence of a coupling between substrate dipolar phonons and electrons in the monolayer.

\section{Summary}
\label{sec.summary}
We have studied the coupling of dipole active (LO) phonons in the depletion regime of SrTiO$_3$ to electrons in a monolayer of FeSe.  LO phonons produce a dipole field which is long ranged, allowing many STO phonon modes to couple to the electrons in the FeSe. However, because the coupling is Coulombic, it is screened by the total charge density fluctuations in FeSe. At the RPA level, we find that the electron fluctuations in FeSe screen most of the electron-phonon interaction, leaving the overall phonon-mediated potential very weak, and unable either to produce a significant contribution to superconductivity or to cause the ``shadow bands'' observed in recent experiments. We therefore conclude that some other (perhaps non-phonon) mechanism is responsible for the observed enhancement of the transition temperature. 

Our calculations were performed within the RPA approximation, which captures the long range Coulomb effects but is not quantitatively accurate in the strongly correlated, low electron density situation relevant to FeSe. It is possible that a more sophisticated calculation including vertex corrections \cite{Yamakawa16} and a better treatment of the electronic screening in the low density limit might change the physics.   Finally, our analyses do not rule out other possibilities that do not couple to the total electron density fluctuations, for example magnetic or nematic fluctuations\cite{Kang16}. 

\section{Acknowledgement}
YZ is grateful to L-F. Arsenault, Z. He, D. Kennes, J. Liu and E. Wilner for helpful discussions.  Computations were carried out on the Yeti High Performance Computing (HPC) Cluster at Columbia University.
YZ is supported by  the Cornell Center for Materials Research with funding from the NSF MRSEC program (DMR-1120296)
AJM is supported by DOE-ER-046169.
We appreciate an anonymous referee for pointing out the error in the original Eq. 25.

\appendix
\renewcommand{\thefigure}{A\arabic{figure}}

\setcounter{figure}{0}

\section{Electrostatic problem and boundary conditions}
\label{A.electrostatics}
We set up the interface normal to $z$ direction. 
STO substrate fills the $z<0$ semi-space with high frequency dielectric constant $\epsilon_\text{STO}$, and the dipoles $\vec{d}$ live in the STO depletion layer that starts at $z=0$ and extends to negative $z_0$, the width of depletion region\cite{yjzhou16}. In general each unit cell contains $N$ frozen dipoles (for STO $N=3$). Each dipole has an effective dipole moment $Z_i$ with $i=1...N$.   
FeSe monolayer is modeled within the $0<z<z_1$ layer, with high frequency dielectric constant $\epsilon_\text{FeSe}$ and a two-dimensional charge density at $z=z_1$, $\rho(\vec{r})=\rho_{2D}(r_{2D})\delta(z-z_1)$. 
$z>z_1$ space is vacuum.
Here for simplicity we assume the high frequency dielectric constant for the space between STO and the FeSe sheet is the same with $\epsilon_\text{FeSe}$. 
In-plane Fourier transform a three dimensional Coulomb potential leaves the out-of-plane dimension in a exponential factor $\exp(-qz)$.

The static Hamiltonian is
\begin{eqnarray}
\label{Ecoul}
H_\text{Coul} &=& \int \frac{d^3r}{a^3} \rho(r) \Phi^e(r) + \rho(r)\Phi^d (r) \nonumber\\
&&+ \sum_i Z_i \vec{d}_i(r) \cdot \nabla \Phi^e(r)+\sum_iZ_i \vec{d}_i(r)\cdot \nabla\Phi^d(r)\nonumber\\
\end{eqnarray}
Here we define the charge density and dipole density with respect to the lattice constant of STO $a$, and thus there are two Fe per unitcell.
$\Phi^e$ and $\Phi^d$ denote the effective potential generated by electrons and dipoles, respectively.

So for the effective potential of electrons, we can assume
\begin{equation}
\Phi^e=
\begin{cases}
\Phi_3^e e^{-q(z-z_1)},& z>z_1 \\
\Phi_2^{e+} e^{q(z-z_1)} + \Phi_2^{e-} e^{-q(z-z_1)}, & 0<z<z_1\\
\Phi_1^e e^{q(z-z_1)},& z<0 
\end{cases}
\end{equation}
with $\Phi_i$ to be determined by boundary conditions.

At  $z=z_1$,
we have
\begin{eqnarray}
&&\Phi_3^e = \Phi_2^{e+} + \Phi_2^{e-} \\
&&q\Phi_3^e + \epsilon_\text{FeSe}q (\Phi_2^{e+} -\Phi_2^{e-}) = 4\pi\rho_q
\end{eqnarray}
At  $z=0$,
\begin{eqnarray}
&&\Phi_1^e e^{-qz_1} = \Phi_2^{e+} e^{-qz_1} + \Phi_2^{e-} e^{qz_1} \\
&&-q\Phi_1^e\epsilon_\text{STO} e^{-qz_1} = -q\epsilon_\text{FeSe} 
(\Phi_2^{e+} e^{-qz_1} - \Phi_2^{e-} e^{qz_1})
\end{eqnarray}
Solving equations above gives
\begin{equation}
\label{cases.phie}
\Phi^e =
\begin{cases}
 \left( 1-e^{-2qz_1} \frac{\epsilon_\text{STO} - \epsilon_\text{FeSe}}{\epsilon_\text{STO} + \epsilon_\text{FeSe}}\right) e^{-q(z-z_1)}  \Phi^e_{2+} , & z>z_1\\
\left[ e^{q(z-z_1)} - e^{-q(z+z_1)}\frac{\epsilon_\text{STO} - \epsilon_\text{FeSe}}{\epsilon_\text{STO} + \epsilon_\text{FeSe}} \right ] \Phi^e_{2+}& 0<z<z_1\\
\frac{2\epsilon_\text{FeSe}}{\epsilon_\text{STO} + \epsilon_\text{FeSe}}e^{q(z-z_1)} \Phi^e_{2+} , & z<0
\end{cases}
\end{equation}
with 
\begin{equation}
\label{phi2e}
\Phi^e_{2+} = \frac{4\pi\rho_q}{q} 
\left[ (\epsilon_\text{FeSe}+1) + (\epsilon_\text{FeSe}-1) \frac{\epsilon_\text{STO} - \epsilon_\text{FeSe}}{\epsilon_\text{STO} + \epsilon_\text{FeSe}}e^{-2qz_1} \right]^{-1}
\end{equation}

Similarly, in-plane Fourier transform the dipole field generated by the semi-infinite space of STO also gives $\exp(-qz)$ factor for the out-of-plane dimension.
We write
\begin{equation}
\Phi^d=
\begin{cases}
\Phi_3^d e^{-qz},& z>z_1 \\
\Phi_2^{d+} e^{qz} + \Phi_2^{d-} e^{-qz}, & 0<z<z_1\\
\Phi_1^d e^{qz},& z<0 
\end{cases}
\end{equation}
with boundary conditions at $z=z_1$,
\begin{eqnarray}
&&\Phi_3^d e^{-qz_1}= \Phi_2^{d+} e^{qz_1}+ \Phi_2^{d-}e^{-qz_1} \\
&&-q\Phi_3^de^{-qz_1} = \epsilon_\text{FeSe}q (\Phi_2^{d+} e^{qz_1} -\Phi_2^{d-} e^{-qz_1})
\end{eqnarray}
and at $z=0$,
\begin{eqnarray}
&&\Phi_2^{d+} + \Phi_2^{d-} =\Phi_1^d \\
&&-q\epsilon_\text{FeSe}(\Phi_2^{d+}-\Phi_2^{d-}) +q\epsilon_\text{STO} \Phi_1^{d}  = \sum_i Z_i d^i_q
\end{eqnarray}

These lead to
\begin{equation}
\label{cases.phid}
\Phi^d=
\begin{cases}
\frac{2\epsilon_\text{FeSe} }{\epsilon_\text{FeSe} + 1} \Phi^d_{2-}e^{-qz}, & z>z_1 \\
\left(e^{-qz} + \frac{\epsilon_\text{FeSe} - 1}{\epsilon_\text{FeSe} +1}e^{q(z-2z_1)} \right) \Phi_{2-}^d, &0<z<z_1 \\
\left( 1+ \frac{\epsilon_\text{FeSe} - 1}{\epsilon_\text{FeSe} +1}e^{-2qz_1} \right)\Phi^d_{2-}e^{qz}, &z<0\\
\end{cases}
\end{equation}
with
\begin{eqnarray}
&&\Phi^d_{2-} = \frac{4\pi\sum_i Z_i e d_i(q,0)}{q} \times\nonumber \\
&&
\left[ (\epsilon_\text{STO} + \epsilon_\text{FeSe}) + (\epsilon_\text{STO} - \epsilon_\text{FeSe})
\frac{\epsilon_\text{FeSe} - 1}{\epsilon_\text{FeSe} +1}e^{-2qz_1} \right]^{-1}
\end{eqnarray}

From the above electric potentials, we can write the interaction terms as 
\begin{eqnarray}
H_{e-d} &=& \int \frac{d^2 q}{(2\pi)^2} \frac{2\pi e^2 \sum_i Z_i \rho_q d^i_{-q} e^{-qz_1}}{q \epsilon_\star} \\
H_{d-e} &=& \int \frac{d^2 q}{(2\pi)^2} \frac{2\pi e^2 \sum_i Z_i \rho_{-q} d^i_{q} e^{-qz_1}}{q \epsilon_\star}
\end{eqnarray}
where the interactions have been summed over $z$, with $q$ here in-plane. 
$\epsilon_\star$ expresses the effective dielectric constant associated to the interacting electron-dipole fields,
\begin{eqnarray}
\epsilon_\star &=& \frac{1}{4\epsilon_\text{FeSe}}
\Big[ (\epsilon_\text{STO}+ \epsilon_\text{FeSe}) (\epsilon_\text{FeSe} +1) \nonumber \\
 && +(\epsilon_\text{STO}- \epsilon_\text{FeSe}) (\epsilon_\text{FeSe} -1)e^{-2qz_1} \Big] 
\end{eqnarray}
This value varies between $(\epsilon_\text{STO} +1)/2$, for $qz_1 = 0$ where the case reduces to the surface of a dielectrics,
 and $(\epsilon_\text{STO}+ \epsilon_\text{FeSe}) (\epsilon_\text{FeSe} +1) /4\epsilon_\text{FeSe}$ for $qz_1 \rightarrow \infty$.

Thus in reciprocal space we may write the static Hamiltonian,
\begin{eqnarray}
H_\text{Coul} &=&\frac{1}{2}\int \frac{d^3 q}{(2\pi)^3}\frac{4\pi e^2}{\epsilon_1  q^2}\rho_q\rho_{-q}+ \nonumber \\
&& iq \frac{4\pi e^2 e^{-q_\parallel z_1}}{q^2\epsilon_\star}\sum_m \left( Z_m d^m_{-q}\rho_{q} - Z_m d^m_{q}\rho_{-q}\right) \nonumber \\
&&+ \sum\frac{4\pi Z_i Z_j e^2}{a\epsilon_2} d^i_qd^j_{-q}
\label{Eq.Hstatic}
\end{eqnarray}
where $q$ is a 3D momentum and $q_\parallel$ is the in-plane component.  $\epsilon_1$ and $\epsilon_2$ represent the effective dielectric functions
\begin{equation}
\epsilon_1=\epsilon_\star \left(\frac{2\epsilon_\text{FeSe}}{\epsilon_\text{STO} + \epsilon_\text{FeSe}}\right)
\left( 1-e^{-2qz_1} \frac{\epsilon_\text{STO} - \epsilon_\text{FeSe}}{\epsilon_\text{STO}+\epsilon_\text{FeSe}}\right)^{-1}
\end{equation}

\begin{equation}
\epsilon_2=\epsilon_\star\left( \frac{2\epsilon_\text{FeSe}}{1 + \epsilon_\text{FeSe}}\right)
\left( 1+e^{-2qz_1} \frac{\epsilon_\text{FeSe} - 1}{\epsilon_\text{FeSe}+1}\right)^{-1}
\end{equation}
For $qz_1 = 0$ $\epsilon_1$ and $\epsilon_2$ reduce to $(\epsilon_\text{STO} +1)/2$, resemble to the dielectric constant at the interface of vacuum and a semi-infinite STO. 
For $qz_1\rightarrow \infty$, they reduce to $(\epsilon_\text{STO}+1)/2$ and $(\epsilon_\text{FeSe}+\epsilon_\text{STO})/2$, respectively.

\section{Decoupling the phonons}
\label{A.decouple}
Next we include the energy of multiple dipole oscillations that contribute to the dynamic energy
\begin{equation}
H_d=\frac{1}{2}\int \frac{d^3r}{a^3}\sum_{ij} K_{ij}d_i(r)d_j(r)+M_{ij}\dot{d}_i(r)\dot{d}_j(r)
\label{Hdipole}
\end{equation}
with $K$  and $M$ the force and mass matrices for transverse optical phonons that are nonpolar. The lowest eigenvalue of $K$  goes to zero at the ferroelectric transition. 

Combining $H_{Coul}$ and $H_d$, we have 
\begin{eqnarray}
H&=&\frac{1}{2}\int \frac{d^3 q}{(2\pi)^3}\frac{4\pi e^2}{\epsilon_1 q^2}\rho_q\rho_{-q}+ \nonumber \\
&& iq \frac{4\pi e^2 e^{-q_\parallel z_1}}{q^2\epsilon_\star}\sum_i \left(Z_i d^i_{-q}\rho_{q} -Z_i d^i_{q}\rho_{-q}\right) \nonumber \\
&&+ \sum_{ij}\left(K_{ij}+\frac{4\pi Z_i Z_j e^2}{a\epsilon_2} - M_{ij}\Omega^2\right)d^i_qd^j_{-q}
\label{AEq.Hq}
\end{eqnarray}

To decouple the dynamic phonon term, we shift the dipoles as 
\begin{equation}
d^i_q\rightarrow d^i_q+i\rho_q\frac{4\pi Z_i e^2e^{-q_\parallel z_1}}{
q\epsilon_\star}\mathcal{D}
\label{shift}
\end{equation}
where we  define
\begin{equation}
\mathcal D = \left(K_{ij}+\frac{4\pi Z_i Z_j e^2}{a\epsilon_2}-M_{ij}\Omega^2\right)^{-1}
\end{equation}
$\mathcal D$ can be viewed as the longitudinal phonon propagator in STO, with the dipole-dipole Coulomb energy term leading to the large LO-TO splitting in STO\cite{STO-Born-charge}. 

Summing over $q_z$ using the $q_z=iq$ pole,
the electron density part Hamiltonian becomes
\begin{eqnarray}
\label{Eq.Hel1}
H_{el}&=&\frac{1}{2}\int \frac{d^2q}{(2\pi)^2}
\frac{2\pi e^2}{ q\epsilon_1}\rho_q\rho_{-q} \nonumber \\
&&\left(1-\frac{4\pi e^2e^{-2qz_1}}{\epsilon_2}\sum_{ij}Z_i\mathbf{\mathcal{D}}_{ij}Z_j\right)\nonumber\\
&=&\frac{1}{2}\int \frac{d^2q}{(2\pi)^2}\rho_q\rho_{-q}V_\text{eff}(q,\Omega)
\end{eqnarray}
where the $\parallel$ subscript is dropped since we are now in the 2D momentum space.

\section{Comparing to conventional theory}
\label{A.convention}

\begin{figure}[ht]
 \includegraphics[width=\columnwidth]{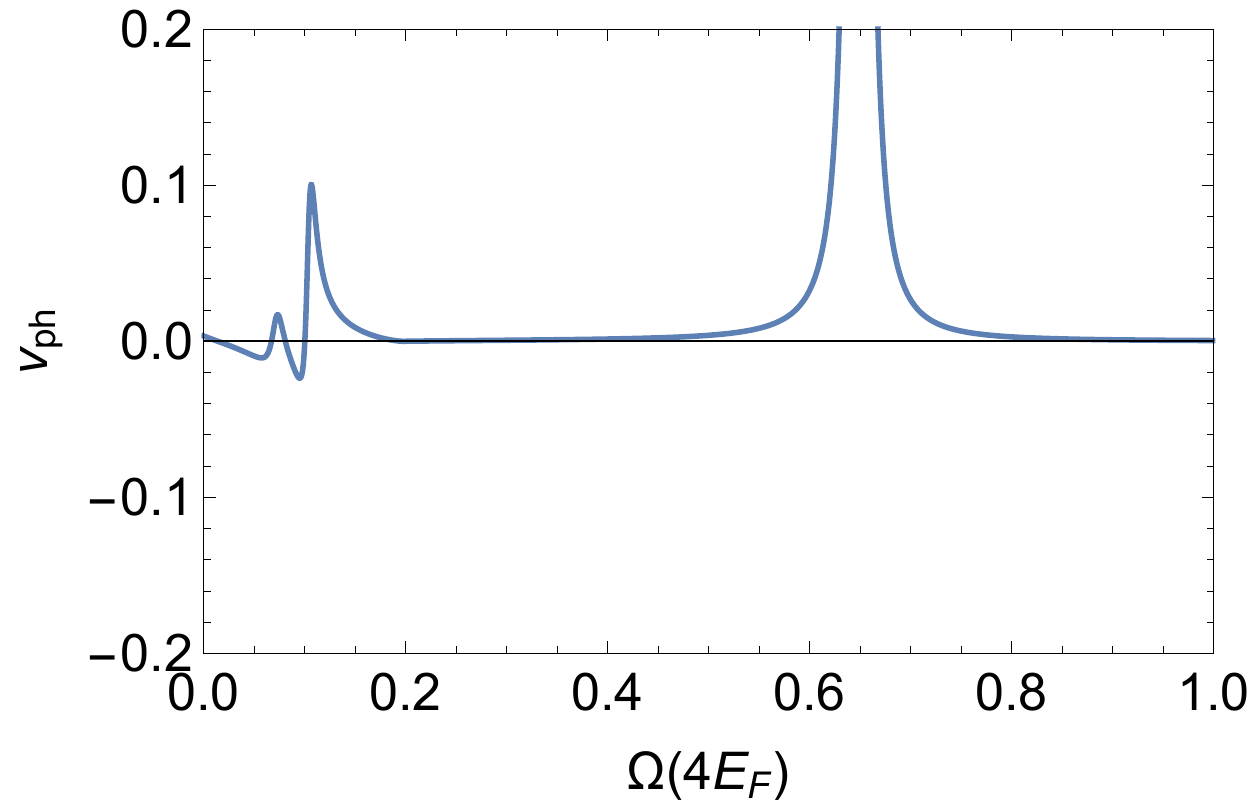}%
 \caption{\label{Fig.vphimag} Imaginary part of $v_{ph}$ in real frequency. $\bar q = 0.2$, $r_S = 5.7$, $z_1 = 1.89$. The strong peak near $\bar \Omega = 0.6$ denotes the plasmon. The peaks near $\bar \Omega = 0.1$ show the poles of phonons.}
  \end{figure}

Comparing to Eq.~(9) in Ref~\onlinecite{Cohen67}, we rewrite Eq~\ref{Eq.Vstar} as 
\begin{eqnarray}
V^\star&=& V_c\frac{1-e^{-2qz_1} g^2 D/V_c}{1-\chi_0 (V_c - e^{-2qz_1} g^2 D)}\nonumber\\
&=&\frac{V_c}{\epsilon_\text{RPA}} -V_c\left[ \epsilon_\text{RPA}^{-1} - \frac{1}{\epsilon_\text{RPA} + \frac{e^{-2qz_1} D}{1- e^{-2qz_1} D}}\right] \nonumber \\
&=& \frac{V_c}{\epsilon_\text{RPA}}\left[ 1 - \frac{D^\star}{\epsilon_\text{RPA}+D^\star}\right]
\label{Eq.Vstar1}
\end{eqnarray}
with the renormalized phonon ``propagator'' and electron-phonon interaction strength $g$,
\begin{equation}
g^2 D = \frac{2\pi e^2}{\epsilon_1 q}\sum_a \frac{\gamma_a\Omega^2_a}{\Omega^2_a + \Omega^2}
\end{equation}
and 
\begin{equation}
D^\star = \frac{e^{-2qz_1} D}{1- e^{-2qz_1} D}
\end{equation}
the dressed phonon ``propagator'' that decay with the set back distance $z_1$, and 
\begin{equation}
\epsilon_\text{RPA} = 1- \chi_0 V_c
\end{equation}
the electronic part of the dielectric constant.

The first term in Eq.~\ref{Eq.Vstar1} gives the RPA-screened Coulomb interaction, 
$V_\text{RPA}=V_c/\epsilon_\text{RPA}$, while the second term contains the screened phonon-induced attractive interaction that may lead to superconductivity. 

The $v_{ph}$ in the main text is exactly the dimensionless form of the second term. 
However, we have to note that $v_{ph}$ should not be thought as conventional propagators. Fig.~\ref{Fig.vphimag} shows the imaginary part of $v_{ph}$ in real frequency. Besides the plasmon peak, the imaginary part changes sign near the poles of LO phonons. The existence of this effect is independent of wavevector.

\bibliography{screening}
\end{document}